\documentclass[useAMS,usenatbib]{mn2e}
\usepackage{amssymb,amsmath,latexsym,graphicx,natbib,mathrsfs}

\usepackage[usenames,dvipsnames]{color}

\title[The formation of RBs via AIC]
{Formation of redbacks via accretion induced collapse}

\author[S. L. Smedley, C. A. Tout, L. Ferrario and D. T. Wickramasinghe]
{Sarah L. Smedley$^1$, Christopher A. Tout$^{1,2,3}$, Lilia Ferrario$^3$\\
\newauthor
and Dayal T. Wickramasinghe$^3$\\
$^1$ Institute of Astronomy, The Observatories, Madingley Road, Cambridge, CB3 0HA\\
$^2$ Monash Centre for Astrophysics, School of Mathematics, Building 28,
Monash University, Clayton, VIC 3800, Australia\\
$^3$ Mathematical Sciences Institute, The Australian National University,
ACT 0200, Australia\\
}

\begin{document}

\date{Accepted 2014 October 24. Received 2014 October 24; in original form 2014 September 03}
\pagerange{\pageref{firstpage}--\pageref{lastpage}} \pubyear{}

\maketitle

\label{firstpage}

\begin{abstract}
We examine the growing class of binary millisecond pulsars known as redbacks.  In these systems the pulsar's companion has a mass between 0.1 and about~$0.5\,\rm M_\odot$ in an orbital period of less than $1.5\,$d. All show extended radio eclipses associated with circumbinary material. They do not lie on the period--companion mass relation expected from the canonical intermediate-mass X-ray binary evolution in which the companion filled its Roche lobe as a red giant and has now lost its envelope and cooled as a white dwarf.  The redbacks lie closer to, but usually at higher period than, the period--companion mass relation followed by cataclysmic variables and low-mass X-ray binaries.  In order to turn on as a pulsar mass accretion on to a neutron star must be sufficiently weak, considerably weaker than expected in systems with low-mass main-sequence companions driven together by magnetic braking or gravitational radiation.  If a neutron star is formed by accretion induced collapse of a white dwarf as it approaches the Chandrasekhar limit some baryonic mass is abruptly lost to its binding energy so that its effective gravitational mass falls.  We propose that redbacks form when accretion induced collapse of a white dwarf takes place during cataclysmic variable binary evolution because the loss of gravitational mass makes the orbit expand suddenly so that the companion no longer fills its Roche lobe.  Once activated, the pulsar can ablate its companion and so further expand the orbit and also account for the extended eclipses in the radio emission of the pulsar that are characteristic of these systems. The whole period--companion mass space occupied by the redbacks can be populated in this way.  
\end{abstract}

\begin{keywords}
stars: neutron - stars: mass-loss - stars: evolution - pulsars:
general - binaries: close
\end{keywords}

\section{Redbacks among millisecond pulsars}
\begin{figure}
\centering
\includegraphics[width=8.9cm]{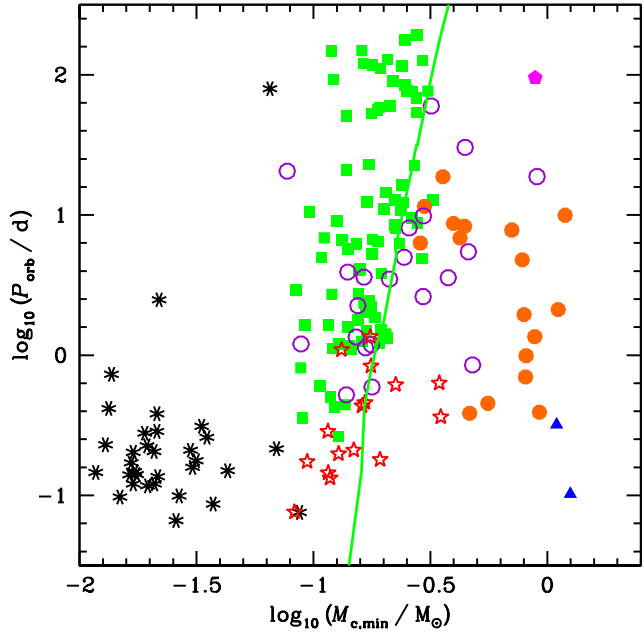}
\caption{Plot of the binary MSPs in the $M_{\rm c}$--$P_{\rm orb}$ plane. The MSP systems have been divided according to companion type, UL (black asterisks), He WD (green squares), C/O WD (orange circles), NS (blue triangles), MS (pink pentagons) and unknown (purple open circles). The mass $M_{\rm c, min}$ is the minimum companion mass for an orbital inclination of $i=90^{\circ}$ with a MSP mass of $1.35\,\rm M_\odot$. RBs are added as open red stars. For consistency with the rest of this work we have used a MSP mass of $1.25\,\rm M_\odot$ when calculating $M_{\rm c, min}$ for these systems. The green line is a recent version of the $M_{\rm c}$--$P_{\rm orb}$ relation for MSPs with He WD companions \citep{2014Smedley}.}
\label{msppop} 
\end{figure}

Millisecond pulsars (MSPs) are defined to have spin periods of $30\,\rm ms$ or less. They have weaker magnetic fields than standard radio pulsars by as much as four orders of magnitude. They are typically old and are usually found in binary systems \citep{Lorimer}. A population of 302 MSPs is listed in the Australian Telescope National Facility (ATNF) Pulsar Catalogue \citep{2005Man}\footnote{http://www.atnf.csiro.au/people/pulsar/psrcat/} and 62 per cent of these are in binary systems. Fig.~\ref{msppop} displays the 93 per cent of binary MSPs that have measured mass functions and hence a derived minimum companion mass in the ATNF Pulsar Catalogue. This is a rough pictorial representation of the different populations. The most common companion to a MSP is a helium-white dwarf (He WD). Such systems make up at least 48 per cent of the binary MSP population. At least 21 per cent have ultra light (UL) companions. These include black widow (BW) systems and MSPs hosting planets. A minimum of 10 per cent of binary MSPs have a carbon/oxygen (C/O) WD, 1 per cent host a main-sequence (MS) star and 1 per cent are paired with a neutron star (NS). Only 2 per cent of the population are triple systems. Redbacks (RBs) account for 9 per cent of the binary MSP population with measured mass functions and currently 10 per cent have unidentified companions. The $M_{\rm c}$--$P_{\rm orb}$ relation for MSPs with He WD companions \citep{2014Smedley} is plotted in Fig.~\ref{msppop} for reference\footnote{Note the $M_{\rm c}$--$P_{\rm orb}$ relation used in this work is extended to lower periods than published in \cite{2014Smedley}. This was done by considering systems with lower initial orbital periods and donor masses.}. 

RBs, plotted as open stars in Fig.~\ref{msppop}, are binary MSPs with orbital periods between about $1.8$ and $32.5\,\rm hrs$. Their companions have minimum masses in the range of $0.1$ to $0.5\,\rm M_{\odot}$. Their orbits have low if not zero eccentricity. They are characterised by mass loss in the vicinity of the companion \citep{2013Roberts}. Long eclipses, often covering most of the orbit \citep{2009Archibald,2011Hessels}, are an observable signature of such mass loss. Long eclipses in the pulsar signal imply that eclipsing regions are larger than the Roche lobes of the stars and the pulsar is seen in some portions of the orbit through the ionised plasma. The favoured companion mass-loss mechanism in RBs is ablation. Ablation is the stripping of mass from the companion as a result of bombardment by the energetic wind of the MSP. Ablation causes the companion star to bloat. Many of the known RBs have been found in follow-up radio surveys of Fermi $\gamma$-ray sources though not all RBs have a $\gamma$-ray counterpart.

\subsection{Observations of Redbacks}

\begin{table*}
\centering
\caption{List of known redback pulsars. The first six are isolated and the rest are in globular clusters. For the companion type, MS means main-sequence star and He means helium white dwarf. Systems where the companion has been observed are indicated with a superscript $\bigtriangleup$ and the systems in which only the pulsar has been observed are marked with a superscript $\bigtriangledown$. If a companion type is not known it is indicated by ? and if a companion type is followed by (?) it means that reasoned deductions have been made on the companion type from the literature. \label{redbacktable}}\vspace{5pt}
\begin{tabular}{| l | c | c | c | c | c | r |} 
\hline
Name & $P_{\rm spin} /\rm ms$ & $P_{\rm orb}/\rm hrs$ ($P_{\rm orb}/\rm d$) & $M_{\rm c}/\rm M_{\odot}$ & $\dot{P}/(10^{-20}\,\rm s/\rm s)$ & Companion type & Reference \\
\hline
J1628--3205 & 3.21 & 5.0 (0.21) & 0.16 & $\star$ & ? & \cite{2012Ray} \\ 
J1816+4510 & 3.19 & 8.7 (0.361) & 0.16 & 4.1 & He(?)$^{\bigtriangleup}$& \cite{2012Kaplan} \\ 
J1023+0038 & 1.69 & 4.8 (0.198) & 0.2 & 1.2 & MS(?)$^{\bigtriangleup}$ & \cite{2009Archibald} \\ 
J2215+5135 & 2.61 & 4.2 (0.18) & 0.22 & $\star$ & MS(?)$^{\bigtriangledown}$ & \cite{2011Hessels} \\ 
J1723--2837 & 1.86 & 14.8 (0.615) & 0.24 & 0.75 & MS$^{\bigtriangleup}$ & \cite{2010Crawford} \\ 
J2129--0429 & 7.61 & 15.2 (0.633) & 0.37 & $\star$ & ? & \cite{2011Hessels} \\ 
\hline  
J0024--7204W & 2.35 & 3.2 (0.133) & 0.14 & $\star$ & MS$^{\bigtriangleup}$ & \cite{2000Camilo} \\
J1701--3006B & 3.59 & 3.5 (0.145) & 0.14 & -34.8 & MS(?)$^{\bigtriangleup}$ & \cite{2003Possenti} \\
J1740--5340A & 3.65 & 32.5 (1.354) & 0.22 & 16.8 & MS$^{\bigtriangleup}$ & \cite{2001DAmico} \\
J1748--2446A & 11.56 & 1.8 (0.076) & 0.10 & 10 & ? & \cite{1990Lyne} \\
J1748--2446P & 1.73 & 8.7 (0.363) & 0.44 & $\star$ & MS$^{\bigtriangledown}$ & \cite{2005Ransom} \\
J1748--2446ad & 1.40 & 26.3 (1.094) & 0.16 & -3.4 & ? & \cite{2006Hessels} \\
J1748--2021D & 13.50 & 6.9 (0.286) & 0.14 & 58.7 & ? & \cite{2008Freire} \\
J1824--2452H & 4.63 & 10.4 (0.435) & 0.20 & $\star$ & ? & \cite{2011Bogdanov} \\
J1824--2452I & 3.93 & 11.0 (0.459) & 0.20 & 1000 & MS(?)$^{\bigtriangleup}$ & \cite{2011Bogdanov} \\
J1910--5959A & 3.27 & 20.1 (0.837) & 0.22 & 0.29 & He(?)$^{\bigtriangleup}$ & \cite{2001DAmico} \\
J2140--2310A & 11.02 & 4.4 (0.174) & 0.10 & -5.2 & ? & \cite{2004Ransom} \\
\hline
\end{tabular}
\end{table*}

 Not all companions to the MSPs in known RB systems have yet been closely studied but a few have been looked at in more detail. \cite{2013Roberts} identified the RB systems in the Galactic field. We list these systems in Table~1 together with the possible globular cluster RBs. The globular cluster RBs were selected from the P. Freire's pulsar database\footnote{http://www.naic.edu/~pfreire/GCpsr.html}, the ATNF Pulsar Catalogue and references therein. There are three main companion types associated with RBs in Table~1, main-sequence stars (MS), helium white dwarfs (He) and those that have unidentified companions marked as `?'. We consider each RB system in turn. 
\par J1723--28 was discovered in the Parkes Multibeam Survey \citep{2004Faulkner}. The light curve of  this system shows an eclipse fraction of 15 per cent. This requires an obscuration of twice the Roche lobe size inferred for the companion star. \cite{2013Crawford} identified the companion in the UV, IR and optical archive data and performed new optical photometry. They made a spectroscopic study of the companion and identified it to be a non-degenerate G-type star. 
\par J1740--5340 is another RB system in our Galaxy that is situated in the globular cluster NGC 6397. \cite{2001DAmico2} obtained orbital parameters and commented that the radio signal of the MSP shows signs of frequent interaction with the atmosphere of the companion star. Their investigation indicates that the companion is bloated and could be the star that spun up the MSP in the standard rejuvenation scenario for MSP production. Subsequent analysis was carried out by \cite{2013Mucciarelli} who measured the chemical abundances of the companion star and their results show that the companion is a deeply peeled star. Comparing the object to their stellar models it seems that there is hydrogen-burning CNO-cycle matter on the surface. They suggest that the companion was once a $0.8\,\rm M_{\odot}$ star that has transferred some 70 to 80 per cent of its mass to the pulsar. Both J1732--28 and J1740--5340 are a good indication that the formation of RBs includes a period of RLOF from a low-mass companion. 
\par J0024--7204W is a system in 47 Tucanae. \cite{2002Edmonds} combined Hubble Space Telescope imaging with Chandra X-ray and Parkes radio data to identify a MS companion to the MSP. They arrived at the conclusion that the companion is a main-sequence star because of its position on the colour-magnitude diagram (CMD) and that it has a  mass exceeding $ 0.13\,\rm M_{\odot}$. 
\par J1748--2446P is in the globular cluster Terzan 5 \citep{2005Ransom}. The pulsar signal in this system has been observed to eclipse for more than half of its orbit. The eclipses are irregular and must be caused by a region that is several solar radii in size indicative of a mass outflow. 
\par J1910--5959A is a MSP in NGC 6752 with a companion observed to be a faint, blue star on a CMD between the main-sequence and the carbon oxygen WD cooling track \citep{2006Cocozza}. Observations made with the ESO VLT \citep{2003Ferraro}. \cite{2006Cocozza} use the system's mass ratio and photometric properties of the companion to constrain the pulsar mass to the range $1.2$ -- $1.5\,\rm M_{\odot}$ and its inclination to more than $ 70^{\circ}$. 
\par The companion to J1910--5959A exhibits two phases of brightening of unequal magnitude close to quadrature. This suggests tidal distortion but at present this hypothesis does not appear to be supported by detailed modelling \citep{2006Cocozza}. 
\par J1701--3001B is a system in NGC 6266. The companion to the pulsar is bright, mysteriously red and has an optical variability of about 0.2 magnitudes. From light curve studies the companion appears to be tidally deformed and there is evidence for mass loss in the form of ionised gas from H$\alpha$ studies \citep{2008Cocozza}. Photometry shows the companion to be over-luminous for its $T_{\rm eff} \approx 6,000\,\rm K$. Its luminosity equals that of the turn off luminosity of its cluster but its colour is redder than expected for a MS star. 
\par J1032+0038 was observed to emerge from a period of Roche lobe overflow and is often called the missing link for explaining the connection between low-mass X-ray binaries (LMXBs) and MSPs. From May 2000 to December 2001 the radio source had a blue spectrum with prominent double-peaked emission lines which are characteristic of an accretion disc. Subsequently it has appeared optically as a star with a solar-like spectrum showing that the system has now ceased RLOF and is in a stage where it appears to exhibit the characteristics of a RB. J2215+5135 has a possibly non-degenerate companion that appears similar to J1023+0038 \citep{2011Hessels}. 
\par J1824--2452I is observed to swing back and forth between a rotation-powered radio pulsar and an accretion-powered X-ray pulsar \citep{2013Papitto}. It is situated in the globular cluster M28. The timescale on which the switch between the modes takes place is consistent with a change in the mass transfer rate. This finding supports the theory that, after a Gyr of mass transfer, the accretion rate declines and allows the radio pulsar mechanism to switch on. 
\par Pulsar J1816+4510 was discovered as part of the Green Bank North Celestial Cap Survey in a pointing chosen to follow up of a Fermi $\gamma$-ray source \citep{2012Kaplan}. A  detailed study was carried out on its companion by \cite{2013Kaplan} and their conclusion is that the star has properties of both a He WD and a non-degenerate star but is not exactly identical to either type.
This system shows evidence of ionised gas eclipses similar to what is seen in RB and BW systems with non-degenerate companions. Spectroscopic analysis of the companion star shows it to be a low-gravity, $\log_{10}(g /\rm cm \,\rm s^{-2}) = 4.9 \pm 0.3$, star with similar properties to a white dwarf but with a much larger radius and with abundances of metals up by a factor of about 10. Radial velocity measurements give the mass of the pulsar $M_{\rm psr} \sin^{3}i = 1.84 \pm 0.11\,\rm M_{\odot}$ and the mass of the companion $M_{\rm comp} \sin^{3}i = 0.193 \pm 0.012\,\rm M_{\odot}$. This system is harbouring a very massive neutron star and its evolutionary status may differ from other RBs as we discuss in section 6.

\par
\begin{figure}
\centering 
\includegraphics[width=8.9cm]{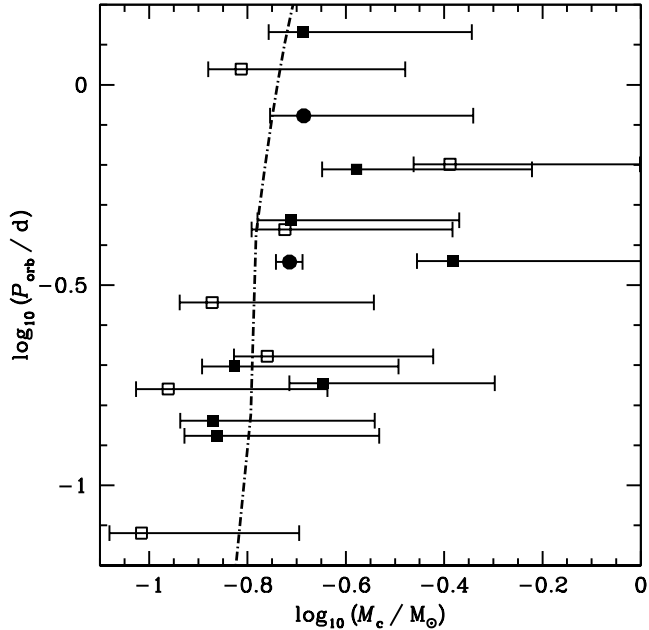} 
\caption{Plot to show where the RB population lies with respect to a recent version of the $M_{\rm c}$--$P_{\rm orb}$ relation (dot-dashed line) for MSPs with He WD companions \citep{2014Smedley}. The systems with MS companions are squares, the He WD companions are circles and the open squares have unknown companions. For the data points the symbols are the median masses ($i = 60^{\circ}$) for a neutron star mass of $1.25\,\rm M_{\odot}$. The minimum mass ($i = 90^{\circ}$) is the lower limit and the maximum mass ($i = 25.8^{\circ}$) is the upper limit for the same neutron star mass.}
\label{He-WD} 
\end{figure}

 We plot these RBs in Fig.~\ref{He-WD} with our \citep{2014Smedley} $M_{\rm c}$--$P_{\rm orb}$ relation for MSPs with He WD companions \citep{1971Refsdal}. 
All of the RBs have measured orbital periods to good precision and minimum companion masses (see these quantities among those listed in Table~1). The calculation of the minimum companion mass depends on the neutron star mass and owing to the absence of measured neutron star masses for RB systems, the two information sources used different neutron star masses in their calculation of this quantity. The analysis of the ATNF Pulsar Catalogue assumes a neutron star mass of $1.35\,\rm M_{\odot}$ while \cite{2013Roberts} used $1.4\,\rm M_{\odot}$. The binary mass function $f$ of each system is
\begin{equation}
f = \frac{(M_{\rm m})^{3}}{(M_{\rm ns} +M_{\rm m})^{2}},
\end{equation}
where $M_{\rm m}$ is the minimum mass and $M_{\rm ns}$ is the neutron star mass. Once we obtained the mass functions we recalculated the minimum mass for a neutron star mass of $M_{\rm ns} = 1.25\,\rm M_{\odot}$. For larger $M_{\rm ns}$ all systems move to the right in Fig.~\ref{He-WD}. To see the range that the companion mass could take we also calculated the median and maximum masses. We calculated all three masses for each system by inverting (by means of interpolation) 
\begin{equation}
f = \frac{(M_{\rm c} \sin i)^{3}}{(M_{\rm c} + M_{\rm ns})^{2}},
\end{equation}
where $M_{\rm c}$ is the mass of the companion star and $i$ is the orbital inclination. The minimum mass is calculated by retrieving the companion mass required for the case where $i = 90^{\circ}$, the median mass by calculating the companion mass required for the case where $i = 60^{\circ}$ and the maximum mass ($90\,$per cent probability limit) with $i = 25.8^{\circ}$. We see from Fig.~\ref{He-WD} that the $M_{\rm c}$--$P_{\rm orb}$ relation for MSPs with He WD companions is close to the RB region but RBs often lie to the right of it. \par To see statistically whether the RBs are consistent with having formed in a similar way to the MSPs with He WD companions, we use the method described by \cite{2014Smedley} to predict the inclinations of the RB systems given the period--companion mass relation and the individual measured binary mass functions for the RBs (calculated from their minimum companion masses). We expect these inclinations to be randomly oriented in space if the RBs are consistent with being formed by the rejuvenation mechanism. We performed a K--S test and the result is that the distribution of the inclination of RBs is distinguishable from random, with a K--S probability of $10^{-3}$ at best for all plausible neutron star masses. Therefore, although the RBs are scattered loosely in the area of the $M_{\rm c}$--$P_{\rm orb}$ relation, the K--S test shows it is rather unlikely the RBs were formed in the same way as the He WD systems.

\subsection{Previous work on RB formation}

\par \cite{2013Chen} studied the formation of both RBs and BWs. They explored the route to these systems from LMXB progenitors with the MESA stellar evolution code \citep{2011Pax}. They found that a factor which determines whether the system evolves to a BW or a RB is the efficiency of the ablation process. They controlled this by an efficiency parameter $f$ and they computed the evolution of binary systems that initially consisted of a $1.4\,\rm M_{\odot}$ NS and a $1\,\rm M_{\odot}$ main-sequence donor in an orbital period of $0.8\,$d. The Roche lobe overflow (RLOF) begins with the companion still on the MS and continues until the donor star becomes fully convective at $M_{\rm d}=0.3\,\rm M_{\odot}$, corresponding to $P_{\rm orb}=2.5\,\rm hr$. At this point they turn off magnetic braking. Interrupting magnetic braking causes the binary star to temporarily detach (cease RLOF). When it detaches the donor is out of thermal equilibrium and inflated following its rapid mass loss episode. During the detached phase it shrinks back into thermal equilibrium. Angular momentum loss by gravitational radiation continues and drives the stars to shorter orbital periods until RLOF resumes. The lack of RLOF for a small range of periods accounts for the period gap in cataclysmic variables \citep{2003Warner}. The top of the gap is when the star detaches and the bottom is when the system reconnects. In the calculations made by \cite{2013Chen} ablation turns on when the system detaches at the top of the period gap and thereafter the evolution is governed by the interplay between ablation and orbital angular momentum loss by the emission of gravitational radiation: ablation increases the period while emission of gravitational radiation decreases it. \cite{2013Chen} propose that the variation in $f$ may be related to the orientation of the magnetic axis with respect to the spin axis so geometric effects are responsible for the two distinct regions that are occupied in the $M_{\rm c}$--$P_{\rm orb}$ plane by RBs and BWs. They note that the RB systems at relatively large companion mass and large orbital periods are not reached by their tracks. \cite{2014Ben} completed a study of the evolution of close binary systems with a radio pulsar member. They considered the scenario in which the systems begin with a NS and a MS donor star in a phase of RLOF. They computed the effect of X-ray irradiation feedback on the mass transfer. They argued that the X-radiation from accretion onto the NS could lead to RLOF occurring in cyclic episodes. There are times when the mass transfer is weak and the pulsar turns on its radio emission mechanism ablating the companion. They conclude that BWs descend from RBs but not all RBs become BWs.

\section{Accretion induced collapse}

\begin{figure}
\centering
\includegraphics[width=5.5cm]{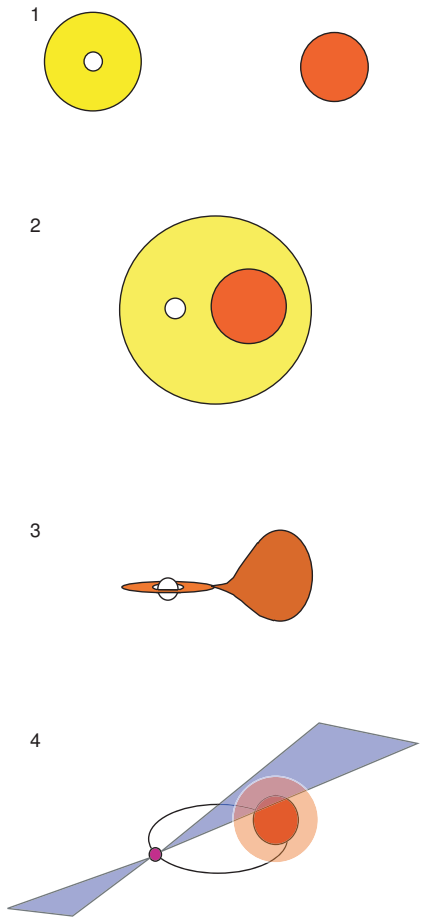}
\caption{Schematic of the stages of the formation of a RB system via AIC. 1) The two companions reside in a wide binary system. The more massive primary star evolves to the SAGB with an ONeMg core. 2) CE phase where the primary core and the secondary spiral towards each other. 3) Magnetic braking brings the stars closer and RLOF commences. 4) AIC occurs forcing detachment of the system and the pulsar radio emission mechanism to initiate. The pulsar then ablates the companion star.}
\label{aic} 
\end{figure}

\par A supernova is not necessary for there to be a pulsar in the system. The production of MSPs by AIC was suggested by \cite{2007Lilia} and extensively investigated by \cite{2010Hurley}. We propose that RBs and some BWs can be formed by accretion induced collapse of an ONeMg WD as a direct result of accretion, without a time lag between the cessation of the mass transfer and the AIC event.
\par If a cool, slowly rotating stellar core composed of oxygen, neon and magnesium (ONeMg) reaches $\rm M_{\rm ec}= 1.37 \rm M_{\odot}$ then electron captures lead to its collapse. There is a loss of pressure support in the star triggered by the sudden capture of electrons by neon and magnesium nuclei. When a stellar envelope remains it is ejected in an electron capture supernova (EC SN). An ONeMg WD that accretes sufficient matter from its companion can similarly collapse. This is accretion induced collapse (AIC). In single stars the zero-age main-sequence mass range that produces ONeMg cores that go through the necessary evolutionary steps to lead to a EC SN is very narrow, only about $0.25\,\rm M_{\odot}$ wide \citep{2006Poe}. Only four per cent of single stars die in this way. In single stars the EC SN occurs when the star is on the super asymptotic giant branch (SAGB, \cite{2010Doherty}). In the most massive AGB stars, core carbon-burning ignites gently before the envelope is lost. Duplicity increases the range over which an EC SN can occur because mass transfer can prematurely remove the stellar envelope. It has been suggested that stars in close binaries systems with masses in the range $8$ to $11\,\rm M_{\odot}$ are likely to collapse in an EC SN, whereas single stars in the same mass range would end there evolution as ONeMg WDs \citep{2004Pod}. Because they lose their hydrogen-rich envelopes via mass transfer before entering the AGB phase they bypass the second dredge-up of elements from their core leaving them with a larger helium core than their single star counterparts. 
\par Initially our systems comprise two stars in a relatively wide detached binary system, the primary with mass between $8$ to $11\,\rm M_{\odot}$ and its companion with a mass of about $1\,\rm M_{\odot}$ (refer to Fig.~\ref{aic}, stage 1). This binary system must be wide enough to allow the core of the primary star to grow an ONeMg core at its centre. This becomes the WD which later collapses to become a MSP. The primary star evolves to overfill its Roche lobe in such a way that a period of common envelope (CE) evolution commences (Fig.~\ref{aic}, stage 2). In this epoch the secondary star and the core of the primary star spiral in towards each other whilst engulfed in the envelope of the primary. The envelope is ejected and the two stars emerge in a now much tighter binary system. The secondary star then evolves to overflow its Roche lobe and starts to transfer matter to its now WD companion, formerly the hot core of the primary star left after the ejection of its envelope. The mass ratio of the companions is such that the RLOF is stable (Fig.~\ref{aic}, stage 3). The WD accretes matter from the secondary star until it reaches $\rm M_{\rm ec} = 1.37 \rm M_{\odot}$. In the AIC the newly formed NS conserves the spin of its progenitor WD but loses some of its gravitational mass to general relativistic binding energy. This mass loss is thought to be around $\Delta M \approx 0.12\,\rm M_{\odot}$ but it depends on the equation of state of the neutron star \citep{2011Bagchi}. It terminates the RLOF because the system widens as the period adjusts to accommodate its new NS inhabitant. Owing to the detachment, the pulsar can now turn on as a radio source and be observed as a MSP. In this phase the system has the characteristics of a RB system (Fig.~\ref{aic}, stage 4). Without any additional mass loss, the orbit of the system in this phase is constantly shortened by the drainage of orbital angular momentum by the emission of gravitational radiation. The orbit would eventually shrink and force another phase of stable RLOF. Such a system would continue to accrete as an X-ray binary. The NS would grow and spin up, leaving the companion further depleted. However radiation from the pulsar can ablate its companion. In this case the orbit grows as the mass of the companion falls because the orbital angular momentum lost from the system is sufficient to avoid further RLOF. \par There are some properties of AIC events that make them attractive for the production of MSPs. When a typical ONeMg WD implodes to form a NS, the conservation of magnetic flux dictates that the final NS should have $B \approx 10^{8}\,\rm G$ which is typical of the magnetic flux of MSPs. Conservation of angular momentum leads to the NS spinning at millisecond periods. The kick of the AIC is thought to be small \citep{2004Pod} so a system in which one star undergoes an AIC has a good chance to remain bound and to have a low spatial velocity. Thus we predict that RBs should have low spatial velocities. There are a couple of RBs listed in the ATNF Pulsar Catalogue with measured proper motions that are consistent with this prediction, but there is not enough data to make a conclusive statistical statement. Note that within the AIC scenario we bypass the need to bury the magnetic field of a normal pulsar during RLOF. \par All measured RB systems have almost circular orbits \citep{2005Man}. During an AIC of a ONeMg WD in a binary system, an orbital eccentricity is induced. In the cases we present in this work the instantaneous expulsion of gravitational mass forces the once circular systems to have about $e=0.2$. This eccentricity is rather short-lived because tidal interaction between the companion star and the newly formed NS quickly circularizes the orbit. The circularization timescale for the binary systems after the AIC is about $10^{4}\,\rm yr$ which is very short compared to the lifetime of our RBs (about $10^{9}\,\rm yr$).\color{black}      

\section{The Code}

We use a version of the Cambridge STARS code \citep{1971Eggleton,1995Pols} updated by \citet{2009Stan}.  The code features a non-Lagrangian mesh. Convection is according to the mixing-length theory of \citet{1958Bohm} with $\alpha_{\rm MLT}=2$ and convective overshooting is included as described by \citet{1997Sch}.  The nuclear species $^{1}\rm H$, $^{3}\rm He$, $^{4}\rm He$, $^{12}\rm C$, $^{14}\rm N$, $^{16}\rm O$ and $^{20}\rm Ne$ are evolved in detail. Opacities are from the OPAL collaboration \citep{1996Iglesias} supplemented with molecular opacities of \citet{1994Alex} and \citet{2005Ferg} at the lowest temperatures and by \citet{1976Buchler} at higher temperatures.  Electron conduction is that of \citet{1969Hub} and \citet{1970Can}.  Nuclear reaction rates are those of \citet{1988Cau} and the NACRE collaboration \citep{1999Angulo}. We include the rate of change of the angular momentum $\dot{J}$ by gravitational radiation \citep{1959Landau},
\begin{equation}
\frac{\dot{J}}{J} = - \frac{32 G^{3}}{5 c^{5}} \left( \frac{M_{\rm a} M_{\rm d} M_{\rm B}}{a^{4}} \right) \quad \mbox{s}^{-1},
\end{equation}
where $M_{\rm a}$ is the mass of the accretor, $M_{\rm d}$ the mass of the donor, $M_{\rm B}$ the mass of the binary system and $a$ the separation of the two stars. We include magnetic braking at the empirical rate of \citet{1981Verbunt},
\begin{equation}
\frac{\dot{J}}{J} = -0.5 \times 10^{-28}\, f_{\rm mb}^{-2} \frac{IR_{\rm d}^{2}}{a^{5}} \frac{GM_{\rm B}^{2}}{M_{\rm a}M_{\rm d}}\quad \rm s^{-1},
\end{equation}
where $R_{\rm d}$ is the radius of the donor star and $I$ its moment of inertia. The factor $f_{\rm mb}$ was chosen to fit the equatorial velocities of G~and K~type stars \citep{smith1979}. 
\par We also include interrupted magnetic braking (IMB) which is common to cataclysmic variables (CVs). IMB is the cessation of magnetic braking when the donor star becomes fully convective. We include this by turning off the magnetic braking when the mass of the convective envelope is $99\,$per cent of the total mass of the star. 

\section{Our Detailed Models of AIC}

We have made models to simulate the evolution of binary systems to, and in some cases through, a RB phase. All of our models begin in a cataclysmic variable stage where a low-mass main-sequence star transfers mass to a WD via RLOF. We are primarily exploring the scenario where RB production is initiated by a detachment of the binary from RLOF as a result of instantaneous gravitational mass loss during an AIC. We engineer the AIC to occur at a various masses $M_{\rm AIC}$ for each system by instantaneously removing an amount of mass whilst conserving the angular momentum of the system. Once the system has detached the radio emission mechanism of the pulsar can operate, ablating the companion and rendering the system observable as an RB. At this stage the binary system exhibits the characteristics of a RB.

\begin{figure}
\centering 
\includegraphics[width=8.9cm]{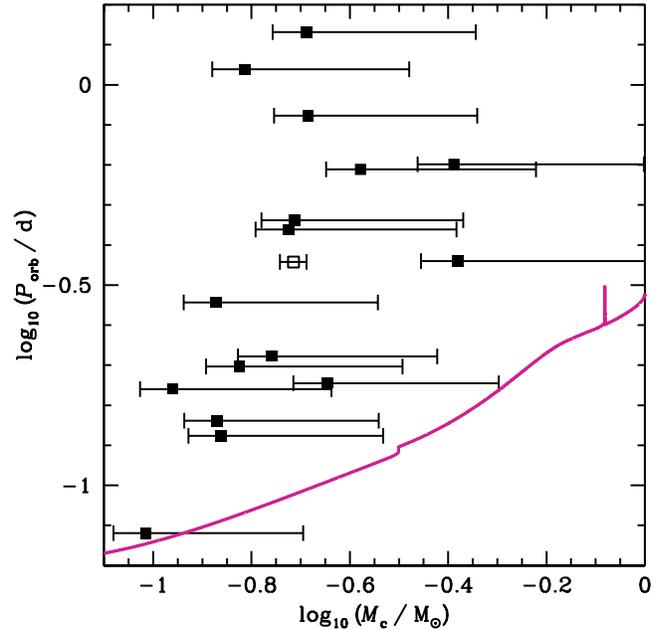} 
\caption{Plot showing a binary system as it evolves through AIC with no subsequent ablation. The initial system has a $1\,\rm M_{\odot}$ donor, a $1.2\,\rm M_{\odot}$ accretor and $P_{\rm orb} = 0.3\,\rm d$. The pink line is the path of the system through the $M_{\rm c}$--$P_{\rm orb}$ plane. The AIC occurs at $\log_{10}(\rm M_{c} / M_{\odot}) = -0.08$ and IBM is at $\log_{10}(\rm M_{c} / M_{\odot}) = -0.50$. The solid squares are the median masses ($i = 60^{\circ}$) of the known redback systems for a neutron star mass of $1.25\,\rm M_{\odot}$. The minimum mass ($i = 90^{\circ}$) is the lower limit and the maximum mass ($i = 25.8^{\circ}$) is the upper limit for a neutron star mass of $1.25\,\rm M_{\odot}$. The open square is J1816+4510 and its observed error bars as given by \citet{2013Kaplan}.}
\label{aicdemo} 
\end{figure}

To demonstrate the main features of our AIC scenario, Fig.~\ref{aicdemo} shows the evolution of such a system that initially contains a $1\,\rm M_{\odot}$ donor and a $1.2\,\rm M_{\odot}$ accretor and in an orbit with a period of $0.3\,\rm d$. The resulting neutron star is $1.25\,\rm M_{\odot}$ with a pre-AIC mass of $1.37\,\rm M_{\odot}$. We chose this initial WD mass of $1.2\,\rm M_{\odot}$ because modelling by \cite{2014Doherty} indicates that this is typical of ONeMg WDs formed in SAGB stars. We chose a $1\,\rm M_{\odot}$ donor as indicative of a low-mass MS star that does not evolve over the timescale of this evolution. Lower-mass donors are less likely to include an AIC because less mass can be transferred. The pink line in Fig.~\ref{aicdemo} is the evolution of the system in the $M_{\rm c}$--$P_{\rm orb}$ plane. As $M_{\rm c}$ decreases from $1\,\rm M_{\odot}$ by RLOF, the period decreases too, until $M_{\rm c} = 0.82\,\rm M_{\odot} $ [$\log_{10}(M_{\rm c}/\,\rm M_{\odot}) = -0.08$] when the WD has accreted sufficient material to increase its mass to $1.37\,\rm M_{\odot}$. At this point we take $0.12\,\rm M_{\odot}$ off the accretor instantaneously and adjust the period to accommodate the new lower-mass neutron star. We assume there is no kick and the orbit is rapidly circularized by the tidal dissipation in the red companion. Without ablation, owing to the loss of orbital angular momentum by both gravitational radiation and magnetic braking, the two companions move closer to each other again and resume RLOF (seen in Fig.~\ref{aicdemo}). RLOF then continues until the companion reaches a point where it becomes fully convective and magnetic braking ceases. This perturbs the system and it detaches again. This occurs at $\log_{10}(M_{\rm c}/\,\rm M_{\odot}) = -0.5$. However, the loss of orbital angular momentum by gravitational radiation is still operating so the companion reconnects once more in RLOF. The only places in Fig.~\ref{aicdemo} that the system could appear potentially as a RB are at $\log_{10}(M_{\rm c}/\,\rm M_{\odot}) = -0.08$ for periods between $\log_{10}(P_{\rm orb}/\,\rm d) = -0.60$ and $-0.51$ (at $0.83\,\rm M_{\odot}$ between $0.25$ and $0.31\,\rm d$) and at $\log_{10}(M_{\rm c}/\,\rm M_{\odot}) = -0.5$ in the period gap.
\par Fig.~\ref{aicdemo} is instructive because it demonstrates that ablation is essential to place our models in the period range of the known RBs. The jump in orbital period at the instance of the AIC is proportional to the amount of mass lost from the system during the collapse. Assuming the mass lost is $0.12\,\rm M_{\odot}$, the tracks can only jump up in orbital period by roughly $0.1\,$d. Even if the AIC occurs at the higher masses, this is not enough of a jump up in orbital period to explain a system such as J1740--5340A at $1.354\,$d. It is fair to say that without ablation of the companion the known RBs cannot be explained by this AIC scenario.

\subsection{Ablation}
We include in our code a prescription for the rate of mass loss from the donor star ($-\dot{M_{\rm d}}$) in the evaporating wind driven by pulsar irradiation \citep{1992Stevens}
\begin{equation}
\dot{M_{\rm d}} = - \frac{f}{2v_{\rm d, esc}^{2}}L_{\rm P} \left( \frac{R_{\rm d}}{a} \right)^{2},
\end{equation}
where $L_{\rm P}$ is the pulsar's spin-down luminosity, $v_{\rm d, esc}$ the escape velocity of a thermal wind from the surface of the donor star and $f$ an efficiency parameter. We write $L_{\rm P} = 4 \pi^{2} I \dot{P} / P^{3}$, where $I$ is the moment of inertia of the pulsar, $P$ its spin period and $ \dot{P}$ its spin-down rate. We use typical $P=3\,\rm ms$, $ \dot{P} = 10^{-20}\,\rm s\,\rm s^{-1}$ and $I=10^{45}\,\rm g\,\rm cm^{2}$. 
\begin{figure}
\centering 
\includegraphics[width=8.9cm]{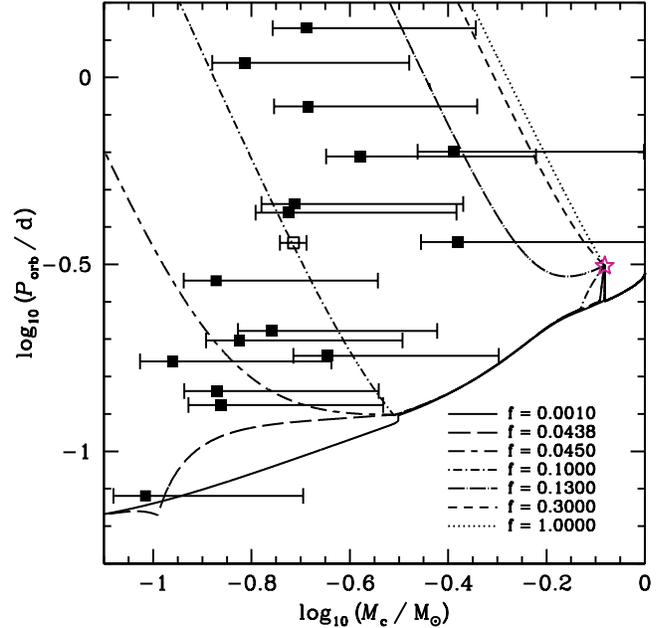} 
\caption{Plot showing the evolution in the $M_{\rm c}$--$P_{\rm orb}$ plane of a binary system with ablation efficiency $f$. The initial system has a $1\,\rm M_{\odot}$ donor, a $1.2\,\rm M_{\odot}$ accretor and $P_{\rm orb} = 0.3\,\rm d$. In all cases the resulting neutron star has a mass of $1.25\,\rm M_{\odot}$ with a pre-AIC mass of $1.37\,\rm M_{\odot}$. The position of the post-AIC system before ablation is marked with a pink star. The solid squares are the median masses ($i = 60^{\circ}$) of the known redback systems for a neutron star mass of $1.25\,\rm M_{\odot}$. The minimum mass ($i = 90^{\circ}$) is the lower limit and the maximum mass ($i = 25.8^{\circ}$) is the upper limit. The open square is J1816+4510 with its observed error bars.}
\label{ablate} 
\end{figure}
\par We made models that included companion ablation after the AIC. Fig.~\ref{ablate} shows the affect of varying the strength of the ablation on the relationship between companion mass and orbital period. There are seven cases plotted for different ablation strengths $f$. All initial systems have a $1\,\rm M_{\odot}$ donor, a $1.2\,\rm M_{\odot}$ accretor, $P_{\rm orb} = 0.3\,\rm d$, resulting neutron star of $1.25\,\rm M_{\odot}$ with a pre-AIC mass of $1.37\,\rm M_{\odot}$. In Fig.~\ref{ablate} we see the two modes of detachment of the binary systems in action. For higher ablation rates of $f \geq 0.12$, ablation is strong enough to overcome the pull of magnetic braking immediately after AIC and the systems evolve to higher periods without reconnecting for a second phase of RLOF. Lower ablation rates of $f < 0.12$, are not strong enough and the systems reconnect for a second phase of RLOF, this time with an accreting NS. For $0.12 > f \geq 0.044$ the ablation rate after detaching at the period gap is strong enough to overcome the angular momentum loss by gravitational wave emission and the systems evolve to higher periods without reconnecting for a third phase of RLOF. For $f < 0.044$ the ablation rate is not strong enough to overcome the angular momentum loss due to gravitational wave emission and the systems reconnect for a third phase of RLOF. Comparing the tracks that our models take through the $M_{\rm c}$--$P_{\rm orb}$ plane with those published by \cite{2013Chen}, they are exhibiting similar behaviour for similar $f$.\color{black} \par Comparing our models to the data, we see that all of the known RB positions in the $M_{\rm c}$--$P_{\rm orb}$ plane can be reached within their error bars when the effect of ablation is included and the strength of the ablation is varied. Although J1816+4510 is in a region of $M_{\rm c}$--$P_{\rm orb}$ space covered by our models, not all of the companion's stellar properties in our model match observations. We discuss this further in section 6. The companions to the MSPs in the RB stage of our models are consistent with being peeled MS stars.

\subsection{Varying the RLOF efficiency}

So far we have ignored the mass loss from the system before the AIC but it is known that accreting WDs in CVs experience nova eruptions in which significant mass may be lost \citep{2003Warner}. To include novae eruptions we allow mass to be lost from the system during mass transfer. Novae outbursts occur when a WD receives H-rich matter via accretion in a close binary system. Over time, the material received by the WD builds up and the bottom of this layer of material is compressed and heated by the material above it owing to the strong surface gravity of the star. Material at the base of the layer becomes electron-degenerate and reaches the conditions required for a thermonuclear runaway during which significant mass may be ejected from the WD in an observable novae outburst \citep{2003Warner}. We explore cases of non-conservative RLOF to include novae in our models. We do this by defining efficiency parameters $\alpha$, the fraction of mass lost by the donor that is transferred to the accretor and $\beta$, the amount of matter actually captured by the accretor over long timescales.  We assume lost material carries away the specific angular momentum of the component from which it is actually lost so that
\begin{equation}
\frac{\dot{J}}{J} = \frac{\dot{M_{\rm d}}}{M_{\rm d}} \left[ \frac{1- \alpha + \alpha (1-\beta) q^{2}}{1-q} \right],
\end{equation}
where the mass ratio $q = M_{\rm d}/M_{\rm a}$. We keep $\alpha = 1$ always based on the assumption that the accretion process is conservative apart from novae eruptions on the WD. We vary $\beta$ to explore the effects of novae eruptions on the systems. 
\begin{figure}
\centering 
\includegraphics[width=8.9cm]{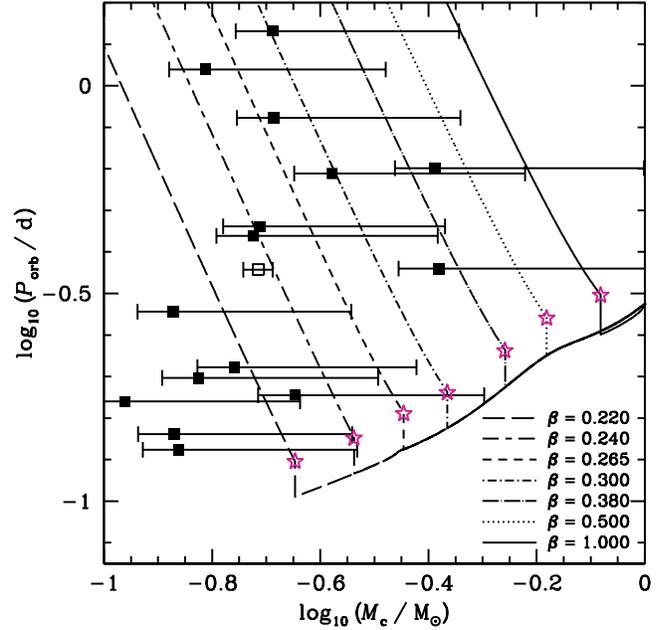} 
\caption{Plot showing the evolution in the $M_{\rm c}$--$P_{\rm orb}$ plane of binary systems with different $\beta$. The initial system has a $1\,\rm M_{\odot}$ donor, a $1.2\,\rm M_{\odot}$ accretor and $P_{\rm orb} = 0.3\,\rm d$. In all cases the resulting neutron star has a mass of $1.25\,\rm M_{\odot}$ with a pre-AIC mass of $1.37\,\rm M_{\odot}$, with the ablation efficiency factor $f=0.45$. The position of each post-AIC system before ablation is marked with a pink star. The squares are the median masses ($i = 60^{\circ}$) of the known redback systems for a neutron star mass of $1.25\,\rm M_{\odot}$. The minimum mass ($i = 90^{\circ}$) is the lower limit and the maximum mass ($i = 25.8^{\circ}$) is the upper limit. The open square is J1816+4510 with its observed error bars. Notice also that the slope of the MSP evolution increases when gravitational radiation alone removes angular momentum. For $\beta < 0.265$ the AIC occurs after interruption of magnetic braking. For larger $\beta$ the AIC occurs first and the change of slope only after the ablation has begun.}
\label{rlof} 
\end{figure}  

\begin{figure}
\centering 
\includegraphics[width=8.9cm]{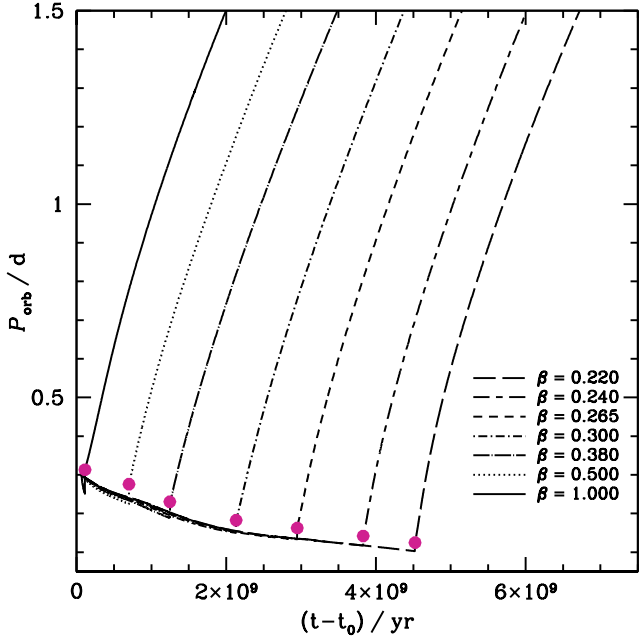} 
\caption{Plot showing the orbital periods of the systems plotted in Fig.~\ref{rlof} against $t-t_{0}$, where $t$ is time in the model and $t_{0}$ is the time when we began our evolution (with a $1\,\rm M_{\odot}$ donor, a $1.2\,\rm M_{\odot}$ accretor and $P_{\rm orb} = 0.3\,\rm d$). The $t_{aic}$ for each model is marked with a circle.}
\label{age} 
\end{figure}

Fig.~\ref{rlof} demonstrates the range of trajectories in the $M_{\rm c}$--$P_{\rm orb}$ plane that the AIC scenario can produce with mass loss in nova eruptions. There are seven different tracks shown in the plot. All have different $\beta$. The initial system has a $1\,\rm M_{\odot}$ donor, a $1.2\,\rm M_{\odot}$ accretor and $P_{\rm orb} = 0.3\,\rm d$. In all cases the resulting neutron star has a mass of $1.25\,\rm M_{\odot}$ with a pre-AIC mass of $1.37\,\rm M_{\odot}$, with the ablation efficiency factor $f=0.45$. For lower $\beta$ less mass is accreted so the AIC occurs at lower companion masses and periods. Note that this lower mass transfer rate means that the donor star is less out of thermal equilibrium than it would be at higher $\beta$ so the period gap is somewhat diminished in the $\beta < 0.265$ cases in Fig.~\ref{rlof}. Cases with $\beta \geq 0.265$ undergo AIC before the period gap. As $\beta \rightarrow 0$ the point of AIC is at ever decreasing companion mass. At $\beta=0.2$ the companions start to reach the lower masses that are associated with BWs as opposed to RBs. In our models, at least 20 per cent of the mass lost by the donor should be retained by the white dwarf for it to undergo an AIC to produce a RB. This requirement is in agreement with the conclusions of \cite{2011Zoro} who claim that WDs should grow in the CV phase to explain the higher mass population of WDs in CVs. \color{black}  We see that the range of orbital periods and companion masses obtained by AIC at points spanning the $M_{\rm c}$--$P_{\rm orb}$ for CVs covers the entire range of the current RB data set in $M_{\rm c}$--$P_{\rm orb}$ space. Our companion star models once again do not match the stellar properties of the companion to J1816+4510.
\par Fig.~\ref{age} demonstrates that the RBs we produce in our models are visible as RBs for a long enough time that they are likely to be observed in this phase. We plot $t-t_{0}$, where $t$ is time of the model and $t_{0}$ is the time when we began our evolution (with a $1\,\rm M_{\odot}$ donor, a $1.2\,\rm M_{\odot}$ accretor and $P_{\rm orb} = 0.3\,\rm d$). We define $t_{\rm aic}$ as the time that the model goes through an AIC (marked as pink circles in Fig.~\ref{age}). The length of time that our RBs are in the region of $M_{\rm c}$--$P_{\rm orb}$ that the known RBs occupy is between $1.6 \times 10^{9}$ and $2 \times 10^{9}\,\rm yr$. 
\par J1816+4510 has a measured mass of $1.84 \pm 0.11\,\rm M_{\odot}$. The masses of post-AIC neutron stars we have presented here are considerably lower than the measured mass of J1816+4510. We investigated the possibility that rapidly rotating white dwarfs avoid AIC until somewhat higher masses and collapse to form more massive neutron stars. \cite{2014Freire} propose the process of super-massive rotating-WD AIC as a direct channel for MSP production. This scenario sees the accreting WD spinning so rapidly that the rate of rotation allows the star to avoid collapse to a neutron star (\citealt{2004Yoon}) even when the mass is well above the Chandrasekhar limit (\citealt{2013Tauris}). An uncertainty lies in the maximum mass of the WD before it collapses to a NS. In the case where the WD is rigidly rotating, the AIC can occur in the range of $1.37$ to $1.48\,\rm M_{\odot}$ \citep{2004Yoon}. For differentially rotating bodies this limiting mass can reach as much as $4\,\rm M_{\odot}$ \citep{2008Piro}. From observation of exceptionally luminous SNe Ia it has been suggested that the pre-SN masses of their WDs were in the range $2$ to $2.5\,\rm M_{\odot}$ (e.g. \citealt{2006Howell}; \citealt{2010Scalzo}). The rotationally-delayed AIC scenario considered by \cite{2014Freire} produces systems that have companion masses between $0.24$ and $0.31\,\rm M_{\odot}$ and orbital periods between 10~and~60$\,\rm d$ because the accretion is by the standard rejuvenation route. RB systems have much shorter periods, despite approximately sharing the same mass range. We made models allowing the AIC to collapse directly at super-Chandrasekhar masses when accreting on our CV scenario. The resulting RBs occupy a similar region in $M_{\rm c}$--$P_{\rm orb}$ space so that a range of neutron star masses from $1.25\,\rm M_{\odot}$ to as much as $2\,\rm M_{\odot}$ could exist in RBs within this scenario though a larger $\beta$, less mass ejected in novae, is required to reach the most massive neutron star masses. We do not find super-Chandrasekhar AICs necessary but equally we cannot exclude them.

\section{The Case of J1816+4510}

The companion to J1816+4510 is intriguing. It displays many of the characteristics of a He WD but with a few key discrepancies. The spectrum of the companion does point towards the conclusion that it is a He WD but is has a surface gravity of $\log_{10}(g/{\rm cm\,s^{-2}})= 4.9 \pm 0.3$ which is lower than a typical He WD, a lower effective temperature of $16{,}000 \pm 500\,$K and a larger radius than expected. For its period it has a larger mass than is predicted by the $M_{\rm c}$--$P_{\rm orb}$ relation for MSPs with He WD companions. Considering the possibility that J1816+4510 formed in the same way as the canonical MSP and He WD detached pair, the lower surface gravity and temperature, along with a larger radius could be consistent with ablation of the companion by the MSP. This process would move the star to lower masses for a given period. However, this would take a system initially on the $M_{\rm c}$--$P_{\rm orb}$ relation for MSPs with He WD companions to the left of it rather than to the right.   
\par In our AIC scenario we can make a model that passes through the data point of J1816+4510 in the $M_{\rm c}$--$P_{\rm orb}$ plane with a mass consistent with observations. It requires near conservative mass transfer to be possible. We begin with an initial WD mass of $1.2\,\rm M_{\odot}$ with a $1\,\rm M_{\odot}$ donor in an orbital period of $0.3\,$d. The WD grows efficiently to $1.37\,\rm M_{\odot}$ then undergoes the AIC losing $0.12\,\rm M_{\odot}$ to become a $1.25\,\rm M_{\odot}$ NS. Ablation then turns on with efficiency $f=0.1$ but this is not strong enough to keep the system detached so there is another phase of RLOF. This RLOF is conservative and the NS grows and spins up until it reaches the period gap, where it detaches as a $1.76\,\rm M_{\odot}$ NS. Ablation turns on again at the same rate of $f=0.1$. This decreases the companion mass and the orbital period. This model's track across the $M_{\rm c}$--$P_{\rm orb}$ plane is the dot-dashed line in Fig.~\ref{ablate}. Not only is this method of obtaining the J1816+4510 system rather finely tuned but the companion does not display all of the same physical properties at the point where the model and the data point coincide. In our model the surface temperature of the companion to J1816+4510 is five times lower than that measured. However the companion in our model is consistent with almost filling its Roche lobe and has a surface gravity of $\log_{10}(g/{\rm cm\,s^{-2}})= 5.01$.
\par To get a He WD in the redback phase, the companion must have evolved off the MS: it must begin RLOF between core hydrogen exhaustion and core helium ignition. In addition the companion must have lost practically all of its envelope in RLOF to reveal its He core in order for the system to be observed as a MSP with a He WD companion. A possible solution to the puzzle of J1816+4510's formation is that the NS was born with a very high mass after an iron-core collapse supernova. In this scenario the pulsar would still need to be spun up by RLOF to millisecond periods via the standard rejuvenation method \citep{1982Alpar,1982Radhak}.   

\section{Conclusion}
We have shown how redbacks can be produced if there is a phase of companion detachment in a close binary in which one component is a rapidly spinning neutron star. Such detachment from Roche lobe overflow interrupts accretion on to the neutron star so that the pulsar's radio emission mechanism can turn on and ablate the companion. We considered two causes of detachment, 1) directly after an accretion induced collapse (AIC) of a ONeMg WD and 2) detachment of an accreting neutron star (formed via AIC) and its companion at the equivalent of the CV period gap when the companion star becomes fully convective. We can split these two causes of detachment into high ablation cases and low ablation case respectively. We explored our AIC scenario by allowing the WD to accrete up to a mass of $1.37\,\rm M_{\odot}$ (from an original mass of $1.2\,\rm M_{\odot}$), when it instantaneously loses $0.12\,\rm M_{\odot}$ of matter to simulate the loss of gravitational mass to internal binding energy of the neutron star. This abrupt mass loss widens the orbit and so terminates the Roche lobe overflow. Thereafter the interplay between ablation and magnetic braking dominates the evolution of the system. Orbital angular momentum loss, owing to the emission of gravitational waves, operates in such systems but it is orbit shrinkage associated with magnetic braking that dominates. In our AIC scenario we see that ablation efficiency governs whether the system remains detached to become a redback or shrinks back to an X-ray binary. For systems with higher ablation rates, of efficiency parameter $f \geq 0.12$, we found that the ablation rate is strong enough to keep the systems detached after an AIC. Lower ablation rates, $f < 0.12$, are not strong enough so such systems enter a second phase of Roche lobe overflow, now with an neutron star accretor. When $0.12 > f \geq 0.044$ the ablation rate is strong enough after detachment at the top of the period gap to maintain a RB state and for $f < 0.044$ the ablation rate is not strong enough and systems reconnect for a third phase of Roche lobe overflow. These systems with very low $f$ could evolve to become black widows. A system only appears as redback when ablation dominates the evolution of the system. By varying the ablation efficiency we can produce models that can populate the redback region in $M_{\rm c}$--$P_{\rm orb}$ space while keeping all Roche lobe overflow wholly conservative. 
\par To investigate the effect that nova eruptions have on redback production we allow for a fraction $(1-\beta)$ of mass accreted by a white dwarf to be expelled with the specific angular momentum of the white dwarf in its orbit. For an initial binary system with a $1\,\rm M_{\odot}$ donor, a $1.2\,\rm M_{\odot}$ accretor and $P_{\rm orb} = 0.3\,\rm d$, we find that varying $\beta$ from 0.2~to~1, whilst maintaining an ablation efficiency of $f=0.45$, thoroughly populates the redback region in the $M_{\rm c}$--$P_{\rm orb}$ plane. The lower $\beta$ the more mass is lost from the system in nova eruptions. For larger $\beta$ we obtain models that reach the higher companion masses and larger periods that are thus far unexplained in literature \citep{2013Chen}.
\par We have compared our models to a list of 17 known redback systems selected from both the ATNF pulsar catalogue and those recorded by \cite{2013Roberts}. We used the minimum mass of each system to obtain its binary mass function and this to calculate the maximum, median and minimum masses for the systems with a neutron star of $1.25\,\rm M_{\odot}$ which is consist with observed AIC remnant masses and projected AIC mass loss from a rigidly rotating white dwarf. We also investigated the possibility that rapidly rotating white dwarfs avoid AIC until somewhat higher masses. These collapse to form more massive neutron stars. The resulting binary MSPs occupy a similar region in $M_{\rm c}$--$P_{\rm orb}$ space so that a range of neutron star masses from $1.25\,\rm M_{\odot}$ to as much as $2\,\rm M_{\odot}$ could exist in RBs within this scenario.
\par Combining the systems that become redbacks straight after AIC with those that become redbacks at the top of the period gap (after an AIC in previous evolution) we conclude that the AIC scenario comprehensively covers the entire RB region of the $M_{\rm c}$--$P_{\rm orb}$ plane. We note that the system, J1816+4510, that has a measured neutron star mass of $1.84 \pm 0.11\,\rm M_{\odot}$ fits in our AIC in many respects but has an observed temperature far higher than the temperatures we are getting in models. We suggest it's previous evolution may be different from the rest of the redback population.

\section*{Acknowledgements}
SLS thanks STFC for her studentship and the Australian National University for hospitality. CAT thanks Churchill College for his fellowship, the Australian National University for a visiting fellowship and the Monash University for support as a Kelvin Watford distinguished visitor. LF and DTW thank the Institute of Astronomy for supporting visits.  


\end{document}